# Light-controlled room temperature ferromagnetism in vanadium-doped tungsten diselenide semiconducting monolayers

Valery Ortiz Jimenez[1], Yen Thi Hai Pham[1], Mingzu Liu[2], Fu Zhang[3,4], Vijaysankar Kalappattil[1], Baleeswaraiah Muchharla[1], Tatiana Eggers[1], Dinh Loc Duong[6,7], Mauricio Terrones[2,3,4,5,*], and Manh-Huong Phan[1,*]

[1]Department of Physics, University of South Florida, Tampa, Florida 33620, USA

[2]Department of Physics, The Pennsylvania State University, University Park, PA 16802 USA

[3]Department of Materials Science and Engineering, The Pennsylvania State University, University Park, PA 16802, USA

[4]Center for Two Dimensional and Layered Materials, The Pennsylvania State University, University Park, PA 16802, USA

[5]Department of Chemistry, The Pennsylvania State University, University Park, PA 16802, USA

[6]Center for Integrated Nanostructure Physics (CINAP) Institute for Basic Science (IBS) Suwon 16419, Republic of Korea

[7]Department of Energy Science, Sungkyunkwan University, Suwon 16419, Republic of Korea.

*Corresponding authors:   phanm@usf.edu, mut11@psu.edu,

**Atomically thin transition metal dichalcogenide (TMD) semiconductors hold enormous potential for modern optoelectronic devices and quantum computing applications. By inducing long-range ferromagnetism (FM) in these semiconductors through the introduction of small amounts of a magnetic dopant, it is possible to extend their potential in emerging spintronic applications. Here, we demonstrate light-mediated, room temperature (RT) FM, in V-doped $WS_2$ (V-$WS_2$) monolayers. We probe this effect using the principle of magnetic LC resonance, which employs a soft ferromagnetic Co-based microwire coil driven near its resonance in the radio frequency (RF) regime. The combination of LC resonance with an extraordinary giant magneto-impedance effect, renders the coil highly sensitive to changes in the magnetic flux through its core. We then place the V-$WS_2$ monolayer at the core of the coil where it is excited with a laser while its change in magnetic permeability is measured. Notably, the magnetic permeability of the monolayer is found to depend on the laser intensity, thus confirming light control of RT magnetism in this two-dimensional (2D) material. Guided by density functional calculations, we attribute this phenomenon to the presence of excess holes in the conduction and valence bands, as well as carriers trapped in the magnetic doping states, which in turn mediates the magnetization of the V-$WS_2$ monolayer. These findings provide a unique route to exploit light-controlled ferromagnetism in low powered 2D spintronic devices capable of operating at RT.**

**Introduction**

Dilute magnetic semiconductors (DMSs) offer an alternative path towards the realization of cutting-edge spintronic devices.[1-6] The use of light to control magnetism in these semiconductors has the added advantage of being able to control both charge and spin simultaneously, which supports the demands of multifunctional (smart) sensing devices, information storage, and quantum computing technologies.[7-10] So far, it has been reported that a carrier-mediated ferromagnetic interaction between the Mn ions in p-type (In,Mn)As/GaSb semiconductor has been enhanced by the illumination of light through the generation of excess holes in the (In,Mn)As layer.[7] Unfortunately, this effect is limited to temperatures (< 50 K), well below ambient temperature, while the most important technological applications are required to operate at room temperature.[11-12]

It has recently been theoretically and experimentally shown that the introduction of a magnetic transition metal atom into semiconducting two-dimensional (2D) TMD, such as V-doped $WSe_2$ and V-doped $WS_2$, permits long-range ferromagnetic order that can be induced at room temperature.[13-15] Currently, the Ruderman–Kittel–Kasuya–Yosida (RKKY) mechanism is believed to be responsible for the long-range ferromagnetic order in these TMD systems, where free holes are the medium that support the interaction between V atoms.[13,14] In particular, we have recently demonstrated that p-type V-doped $WS_2$ monolayers have strong and tunable room temperature ferromagnetism.[15] By replacing W, having six valence electrons, with V, having five valence electrons, an electron deficiency is created in V-$WS_2$ that eventually becomes a p-type dominant semiconductor. Unlike diamagnetic pristine $WS_2$ (Fig. S1(a)), ferromagnetism is enhanced with V doping in monolayers of V-$WS_2$, which is found to be optimized at ~2 at.% V (Fig. S1(b)).[15] The long-range ferromagnetic order in this optimally magnetic 2D DMS allows us to modify its magneto-electronic property with external stimuli, like a magnetic field, an electric field, or as we show in this letter, with light. Photoluminescence (PL) reveal that, even after doping, strong photoluminescence is still present (Fig. S1(c)). These observations lead us to propose that the ferromagnetism in the V-WS$_2$ monolayer can be mediated by illumination with a laser of appropriate energy, that is, above the optical gap (see Fig. 1). Electrons from photogenerated electron-hole pairs may be captured by the V atoms, thus creating an imbalance in the carrier population (i.e., the generation of excess holes) such that the ferromagnetism of the monolayer is modified. While p-type (In,Mn)As/GaSb showed light-mediated ferromagnetism at temperatures below 50 K,[7] in this manuscript we demonstrate that light controls the ferromagnetism at ambient temperature in an atomically thin p-type V-WS$_2$ semiconductor. The light mediated changes of the magnetization in 2D semiconducting materials, will certainly lead to novel applications in spintronic devices that have not been yet realized.

**Probing light-induced 2D magnetism**

Probing magnetism in atomically thin materials is extremely challenging when compared to bulk systems.[16,17] While techniques such as vibrating sample magnetometry (VSM) and superconducting quantum interference devices (SQUID) are capable of measuring the magnetic moment of these materials,[18,19] measurements in real time while applying external stimuli is not easily achievable. Methods that require electrical contact, such as transport measurements, have an extra layer of difficulty since monolayer films often do not span the surface continuously and the size of the electrical contacts are large compared to the surface area.[20] Optical methods based on the magneto-optical Kerr effect (MOKE) have proven very successful in thin films and have had a measure of success with 2D materials such as CrI$_3$.[16] However, local heating from high laser powers causes thermal instability, which is a significant source of noise in these measurements. Therefore, cryogenic temperatures are needed to reduce thermal and mechanical measurement noise, but for room temperature, this is not an option. Despite this, optical measurements remain a powerful practice that yield crucial insight into the spin dynamics of atomically thin materials.[21,22] The shortcomings of these techniques motivate the development of a new approach to measure 2D magnetization in real time as external stimuli are applied.

In order to probe the light-induced magnetization of an atomically thin magnetic film such as V-WS$_2$ monolayers, we propose a new technique utilizing our recently developed magneto-LC resonance sensor with ultrahigh magnetic field sensitivity (pT regime).[23,24] The sensing element of this device is a magnetic microwire wound into a coil driven with a frequency of 118 MHz, which is near the coil's LC resonance (Fig. S2). The impedance of the coil is then measured with an impedance analyzer, from which we extract the reactance of the coil. The setup is depicted in Fig. 1. The film is placed at the core of the coil and the reactance of the coil is measured. Theoretically, the reactance of the coil is written as

$$X_{\text{coil}} = \frac{\omega[L(1 - \omega^2 LC_{par}) - C_{par}R_{par}^2]}{(1 - \omega^2 LC_{par})^2 + (\omega C_{par}R_{par})^2},$$

where $\omega$ is the angular frequency. Since the film is ferromagnetic, it will modify the relative permeability of the space within the coil, thus changing the magnetic flux through the coil and consequently the reactance of the coil. As the microwire itself is ferromagnetic, the magnetization of the film will also lead to a change in the effective permeability of the microwire. Therefore, the reactance of the sensor depends on this effective permeability, $X = X(\mu_{eff})$. Changes in the permeability of the film upon light illumination will influence the effective permeability of the coil, which can be accessed through the change in its reactance, $\Delta X = X(\mu_{eff}, laser\ on) - X(\mu_{eff}, laser\ off)$.

**Results and Discussion**

Due to the presence of defects and edge effects in $WS_2$ monolayers, a small magnetic moment may be present in these "pristine" films. Therefore, we first examine whether photogenerated electron-hole pairs have any effect on the magnetization of the film. Figure 2(a) shows the signal obtained from a $WS_2$ monolayer film with and without laser excitation. In this figure, a small change in the signal, likely due to thermal fluctuations in the film, can be seen. Next, we performed the measurement on the 2 at.% V-doped $WS_2$ monolayer (Fig. 2(b)), where we observe a large change in reactance (magnetic permeability or magnetization) upon illumination with the same laser. The excitation wavelength is $\lambda = 650$ nm ($h\nu = 1.91$ eV) with the power of 5.1 mW/cm$^2$ at the sample surface. The laser was operated for several minutes during these measurements, which were carried out at room temperature. Since prolonged laser exposure may heat up the magnetic coil and hence change the magnetic properties, coil heating effects due to laser exposure have also been studied. As shown in Fig. S3, the dot laser, which covers only a small area of the coil (0.11 cm$^2$), has a negligible effect on the magnetism of the coil. These findings indicate that the observed enhancement of the magnetization/permeability from the experimental setup, i.e. illuminated V-$WS_2$ monolayer (Fig. 2b) within the coil, is not due to a laser/sample heating effect but originates from carrier-mediated ferromagnetism, similar to the case of a p-type (In,Mn)As/GaSb semiconductor.[7] As shown in Fig. S4, measurements on the same V-$WS_2$ sample were performed weeks apart to demonstrated that we can reproduce the change in magnetization, and confirm the reversibility and reproducibility of this effect. Measured magnetic hysteresis (*M-H*) loops at room temperature before and after illumination also confirm that the process is reversible (Fig. S5). It is worth mentioning that upon light illumination with comparable laser powers, the magnetization change in the illuminated (In,Mn)As/GaSb film ($\lambda$ = 685 nm, 6 mW/cm$^2$) was only observed below 50 K, while enhanced magnetization is achieved at room temperature for the illuminated V-WS$_2$ monolayer ($\lambda$ = 650 nm, 5.1 mW/cm$^2$). This striking difference makes this atomically thin ferromagnetic semiconductor a promising candidate for use in light-controlled spintronics and other multifunctional nanodevices. The semiconducting nature of V-WS$_2$ facilitates incorporation to current silicon-based technology and provides a platform for optoelectronic phenomena; combined with its FM properties we obtain a unique way to manipulate the spin states in the material by illuminating it with light.

It is of interest to determine how an increase in illumination area of the sample would affect the permeability/magnetization, so we illuminated the film with two 650 nm lasers, labeled 'dot' and 'target' lasers, with 0.11 cm$^2$ and 0.41 cm$^2$ coverage areas, respectively. The two lasers have different spot sizes, but the same light intensity of 4.2 mW/cm$^2$. In Fig. 3(a,b) it is observed that increasing the coverage area increases the change in magnetization. Finally, we sought to determine the light intensity dependence of the change in magnetization. In Fig. 3(c,d) we demonstrate this for both 650 nm lasers, both of which show a similar trend. Initially, we see a sharp increase in magnetization with increasing light intensity, and at higher laser intensities the change in magnetization begins to saturate. Since higher laser powers bring about a considerable heating effect, which may damage the coil or the sample, in this study we restricted the laser intensity below 6 mW/cm$^2$ for light-induced magnetization experiments. Photon concentration was calculated as a function of laser intensity, for each laser, using the relation $E = nh\nu$ (where $E$ is the energy and $\nu$ is the frequency, $n$ is the photon-generated carrier concentration, and $h$ is Planck's constant), assuming 2% absorption in the V-WS$_2$ layer [25] and assuming that only 1% of the electrons/holes created do not immediately recombine. We observe that by increasing the illumination area, a smaller photon concentration, compared to the "dot" laser, is required to achieve a change on magnetization. A smaller photon concentration is also necessary to achieve the saturation feature; using the "dot laser" a concentration of ~3.1 x $10^{12}$ photons/cm$^2$/s is necessary to approach saturation, while using the "target laser" saturation starts around ~2.7 x $10^{12}$ photons/cm$^2$/s. This points to a long range cumulative effect in which the enhanced magnetic moments in the illuminated area may couple with the moments surrounding it.

To elaborate these findings, the band structures of the WS$_2$ and V-WS$_2$ monolayers were investigated by density functional theory (DFT) calculations and the results are shown in Fig. S6. In order to establish the change of the WS$_2$ band structure by a V atom, the valence band edge was used as the reference. The V atom induces two empty doping states: one is below the conduction band and the other is on the top of the valence band. These two doping bands are flat and localized, implying a weak interaction with W and S atoms. The latter atom plays the role of electron acceptor, where it accepts thermally excited electrons from the valence band and manifests the p-type characteristic of the V-WS$_2$ monolayer.[15] In addition, there are two strong hybridization bands between the V and W atoms, which induce approximately 1μ$_B$ of the magnetic moment in the V-WS$_2$ monolayer. The energy of the ferromagnetic state is 0.24 eV, which is lower than that of the no-spin state. Similar hybridization bands have been reported in monolayers of V-doped WSe$_2$ [13,14], which also shows a lowering of the energy of a FM state with Curie temperature above RT. This points to V-doped WSe$_2$ monolayers as another candidate for light mediated magnetism.

Generally, two main factors have been suggested to influence the magnetic moment of a DMS upon light illumination:[26-30] (i) the population of the free excited carries in the conduction and valence bands[28], and (ii) the localized excited carriers trapped by magnetic doping states in the band gap of the host material.[26] The former effect is usually dominant in lightly doped samples (e.g. 1.1% of Mn in GaAs[28]), whereas the latter becomes significant in heavily doped samples (e.g. 10% of Mn in CdSe and HgTe)[26,27], in which the dopants form a new band inside the gap of the host. Due to its single layer limit and V concentration of approximately 2%, both free and localized excited carriers are expected to mediate the magnetization in the V-WS$_2$ film when illuminated with an appropriate power laser. Figure 4(a) shows the distribution of the magnetic moments under varying carrier populations. Increasing the concentration of holes results in a more robust magnetic moment across the lattice, where W atoms far from the V site show an enhanced magnetic moment. The evolution of the band structure is calculated under different carrier concentrations, as shown in Fig. 4(b). We find that while the Fermi level is shifted deeper inside the valence band with hole injection, it is shifted toward the conduction band edge with electron injection. The evolution of the exchange energy is also calculated, the result of which is presented in Fig. 4(c). The exchange energy becomes stronger with increasing hole concentration, so the magnetic moment must also change as the carrier concentration varies. Indeed, we observe an increase in the magnetic moment with increasing hole concentration (Fig. 4(c)). This is consistent with our experimental findings (Fig. 3c,d), where a higher light intensity resulted in a larger hole concentration and consequently an increased magnetic moment. At large hole concentrations the magnetic moment saturates, confirming the saturation feature we observed experimentally (Fig. 3c,d). This is also consistent with another report,[31] in which hole injection into a V-WSe$_2$ monolayers increase the magnetic moment. We should note that both electrons and holes are populated in experiments whereas the simulation considers the separated effect for each type of carriers.

In summary, we have demonstrated that magnetism can be tuned with light in V-doped WS$_2$ monolayers, by varying the light intensity or by changing the illumination area. As the film is illuminated, the absorbed photons generate electron-hole pairs and the electrons are captured by the electron deficient V sites, which generate an imbalance in the carrier population and, hence, a change in magnetic moment. We have shown that the carrier concentration can be tuned by changing the light intensity, allowing control over the magnetic moment of the film. Density functional calculations confirm that the magnetic moment of the V-WS$_2$ monolayer can be enhanced by increasing the hole concentration. All of this is achieved at room temperature which has been a key obstacle to applied 2D spintronics. These findings highlight the potential for cutting-edge applications of 2D DMS and other atomically thin magnetic semiconductors.

## Methods

*Sample synthesis and characterization:*

The WS$_2$ and 2 at.% V-doped WS$_2$ monolayer films were synthesized using a reliable single-step powder vaporization method. STEM-EDS of the samples were performed by a FEI Talos F200X microscope with a SuperX EDS detector operating at 200 kV. Photoluminescence (PL) spectra of the samples were recorded with a Coherent Innova 70C argon-krypton laser at 532 nm. Details of the samples' synthesis and characterization have been reported previously.[15]

*Magnetic and light-induced measurements:*

Magnetization versus magnetic field (*M-H*) measurements were carried out at room temperature in a Physical Property Measurement System (PPMS) from Quantum Design with a vibrating sample magnetometer (VSM) magnetometer in fields up to 9 T. The *M-H* loops of the V-WS$_2$ monolayer film before and after light illumination ($\lambda$ = 650 nm, 5.1 mW/cm$^2$) were also recorded at room temperature to investigate the magnetization relaxation of the sample.

Light-induced magnetization measurements on the WS$_2$ and V-WS$_2$ monolayer films were performed with a magnetic microwire LC resonator as the sensing element (see Fig. S2).[24] The reactance (*X*) is monitored as the light illuminates the film with an HP 4191A impedance analyzer. Two different lasers of 650 nm wavelength but different spot sizes were used. One, which is referred to as the 'dot laser,' has a round shaped spot and covers an area of 0.11 cm$^2$. The second, which is referred to as the 'target laser', has a target-shaped spot, in which one of the lines covers the entire surface of the film with a surface area of 0.41 cm$^2$. The target laser has a significant heating effect on the coil and demonstrates a reactance change of about $\Delta X = 1 \, \Omega$. To correct for this, we subtracted $\Delta X$ from the data obtained using this laser. To prevent significant heating effects, low powered lasers were used (< 5 mW).

*Theoretical calculations and simulation:*

The band structures of the pristine WS$_2$ and V-doped WS$_2$ monolayers were calculated using Quantum Espresso code.[32] Projector-augmented wave potentials were used with a cut-off energy of 30 Ry.[33] A 8x8x1 supercell, which corresponds to 1.6% of V atoms, was used with a 3x3x1 k-point grid. The structures of WS$_2$ and V-doped WS$_2$ were optimized until the convergence of force and energy is smaller than 0.0001 Ha and 0.001 (Ha/bohr), respectively, without including spin-orbit coupling. The initial spin state was induced along *z* direction and the total energy is optimized with 10$^{-6}$ Ha of convergence including spin-orbit coupling. To describe the strong correlation of the *d* electrons, the GGA+U method with U = 3 was used for the vanadium atoms.

## Acknowledgments

Research at USF was supported by the U.S. Department of Energy, Office of Basic Energy Sciences, Division of Materials Sciences and Engineering under Award No. DE-FG02-07ER46438. Research at PSU was supported by the Air Force Office of Scientific Research (AFOSR) through grant No. FA9550-18-1-0072 and the NSF-IUCRC Center for Atomically Thin Multifunctional Coatings (ATOMIC).

## Author Contributions

V.O.J. and M.H.P. conceived the initial concept. V.O.J. performed light-controlled magnetization experiments and analyzed the data with inputs from T.E. and B.M. Y.H.T.N. and V.K. performed magnetic measurements and analyzed the magnetic data. F.Z. and M.L. synthesized the films and characterized the structural and optical properties. L.D.D. performed the computational calculations. V.O.J. wrote the manuscript with inputs from other authors. M.H.P. directed the research.

## Competing interests

The authors declare no competing financial interest.

# References

[1] Dietl, T. A ten-year perspective on dilute magnetic semiconductors and oxides. *Nat. Mater.* **9**, 965-974 (2010)

[2] Dietl, T., Bonanni, A. & Ohno, H. Families of magnetic semiconductors – an overview. *Journal of Semiconductors* **40**, 8 (2019)

[3] Takiguchi, K., Anh, L. D., Chiba, T., Koyoma, T., Chiba, D. & Tanaka, M. Gian gate-controlled proximity magnetoresistance in semiconductor-based ferromagnetic-non-magnetic bilayers. *Nat. Phys.* **15**, 1134-1139 (2019)

[4] Wang, J.Y., Verzhbitskiy, I., & Eda, G. Electroluminescent Devices Based on 2D Semiconducting Metal Dichalcogenides. *Adv. Mater.* **30**, 1802687 (2018)

[5] Yi, Y., Chen, Z., Yu, X.F., Zhou, Z.K., & Li, J. Recent Advances in Quantum Effects of 2D Materials. *Adv. Quantum Technol.* **2**, 1800111 (2019)

[6] Kong, T., et. al. $VI_3$–a New Layered Ferromagnetic Semiconductor. *Adv. Mater.* **31**, 1808074 (2019)

[7] Munekata, H. & Abe, T. Light-induced ferromagnetism in III-V-based diluted magnetic semiconductor heterostructures. *J. Appl. Phys.* **81**, 4862 (1997)

[8] Shankar, H., Sharma, A. & Sharma, M. Optically Induced Ferromagnetism in III-V Dilute Magnetic Semiconductors. *Integrated Ferroelectrics* **203:1**, 67-73 (2019)

[9] Mishra, S. & Satpathy, S. Photoinduced Magnetism in the Ferromagnetic Semiconductors. *International Journal of Modern Physics B* **Vol. 24**, No. 03, 359-367 (2010)

[10] Zhu, L., et. al. Photoinduced magnetization effect in a p-type $Hg_{1-x}Mn_x Te$ single crystal investigated by infrared photoluminescence. *Phys. Rev. B* **94**, 155201 (2016)

[11] More than just room temperature. *Nat. Mater.* **9**, 951 (2010)

[12] Coelho, P. M., et. al. Room-Temperature Ferromagnetism in MoTe$_2$ by Post-Growth Incorporation of Vanadium Impurities. *Adv. Electron. Mater.* **5**, 1900044 (2019)

[13] Duong, D.L., Yun, S.J., Kim, Y.K., Kim, S.G., & Lee, Y.H. Long-range ferromagnetic ordering in vanadium-doped WSe$_2$ semiconductor. *Appl. Phys. Lett.* **115**, 242406 (2019)

[14] Yun, S.J., et. al. Ferromagnetic Order at Toom Temperature in Monolayer WSe$_2$ Semiconductor via Vanadium Dopant. *Adv. Sci.* **7**, 1903076 (2020)

[15] Zhang, F., et. al. Monolayer Vanadium-doped Tungsten Disulfide: An Emerging Room-Temperature Diluted Magnetic Semiconductor. https://arxiv.org/abs/2005.01965

[16] Huang, B., et. al. Layer-dependent ferromagnetism in a van der Waals crystal down to the monolayer limit. *Nature* **546**, 270-273 (2017)

[17] Gong, C., et. Al. Discovery of intrinsic ferromagnetism in two-dimensional van der Waals crystals. *Nature* **546**, 265-269 (2017)

[18] Bonilla, M. Strong room-temperature ferromagnetism in VSe$_2$ monolayers on van der Waals substrates. *Nature Nanotech.* **13**, 289-293 (2018)

[19] O'Hara, D.J., et. al. Room Temperature Intrinsic Ferromagnetism in Epitaxial Manganese Selenide Films in the Monolayer Limit. *Nano. Lett.* **18**, 5, 3125-3131 (2018)

[20] Bushong, E.J. et. al. Imaging Spin Dynamics in Monolayer WS$_2$ by Time-Resolved Kerr Rotation Microscopy. arXiv:1602.03568 [cond-mat] (2016)

[21] Aivazian, G. et. al. Magnetic Control of Valley Pseudospin in Monolayer WSe$_2$. *Nat. Phys.* **11**, 148-152 (2015)

[22] Avsar, A., et. al. Defect induced, layer-modulated magnetism in ultrathin metallic PtSe$_2$. *Nature Nanotech.* **14**, 674-678 (2019)

[23] Thiabgoh, O., Eggers, T., & Phan, M.H. A new contactless magneto-LC resonance technology for real-time respiratory motion monitoring. *Sensor Actuat. A-Phys.* **265**, 120-126 (2017)

[24] Ortiz Jimenez, V., et. al. A magnetic sensor using a 2D van der Waals ferromagnetic material. *Sci. Rep.* **10**, 4789 (2020)

[25] Zhu, B., Chen, Xi. and Cui, X. Exciton Binding Energy of Monolayer $WS_2$. *Sci. Rep.* **5**, 9218 (2015)

[26] Warnock, J., et. al. Optical orientation of excitons in (Cd, Mn)Se and (Cd,Mn)Te. *Solid State Commun.* **54**, 215-219 (1985)

[27] Krenn, H., Zawadzki, W., and Bauer, G. Optically Induced Magnetization in a Dilute Magnetic Semiconductor $Hg_{1-x}Mn_xTe$. *Phys. Rev. Lett.* **55**, 1510 (1985)

[28] Oiwa, A., Mitsumori, Y., Moriya, R., Slupinski, and Munekata, H. Effect of Optical Spin Injection on Ferromagnetically Coupled Mn Spins in the III-V Magnetic Alloy Semiconductor (Ga, Mn)As. *Phys. Rev. Lett.* **88**, 137202 (2002)

[29] Sun, B., et. al. Photoinduced spin alignment of the magnetic ions in (Ga,Mn)As. *J. Appl. Phys.* **100**, 083104 (2006)

[30] Jungwirth, T., et. al. Spin-dependent phenomena and device concepts explored in (Ga,Mn)As. *Rev. Mod. Phys.* **86**, 855 (2014)

[31] Duong, D.L., Kim, S.G. & Lee, Y.H. Gate modulation of the long-range magnetic order in a vanadium-doped $WSe_2$ semiconductor. *AIP Advances* **10**, 065220 (2020)

[32] Giannozzi, P., *et al.*, QUANTUM ESPRESSO: a modular and open-source software project for quantum simulations of materials, *J. Phys. Condens. Matter* **21**, 395502 (2009)

[33] E. Kucukbenli, M. Monni, B. I. Adetunji, X. Ge, G. a Adebayo, N. Marzari, S. de Gironcoli, and a D. Corso, http://Arxiv.Org/Abs/1404.3015 (2014).

**Figure captions**

**Fig. 1: (a)** An illustration of the measurement scheme for light-induced magnetization of the V-WS$_2$ monolayer; **(b)** a sketch showing photon absorption generating a hole-electron pair in the film.

**Fig. 2:** The change in reactance upon illumination with a 650 nm laser for **(a)** WS$_2$ and **(b)** V-WS$_2$ as measured by the microwire coil sensor. The change in reactance is proportional to the change in magnetization. A negligible change in magnetization is observed in pristine WS$_2$ while a significant change in magnetization was measured on the V- doped WS$_2$ monolayer.

**Fig. 3:** The dependence of the change in reactance ($\Delta X$) on illumination area is shown in **(a,b)**; **(c,d)** shows the dependence of $\Delta X$ on light intensity and photon concentration for two different illumination areas. All measurements shown were made using a 650 nm light sources, labeled "dot" and "target lasers with illumination areas 0.11 cm$^2$ and 0.41 cm$^2$, respectively. We observed a similar trend for both lasers in which increasing light intensity (and photon concentration) results in an increase in magnetization, until it reaches a critical photon concentration at which saturation begins. Notably, a smaller photon concentration (~2.7 x 10$^{12}$ photons/cm$^2$/s) is required to begin saturation using the "target" laser than with the "dot" laser (~3.1 x 10$^{12}$ photons/cm$^2$/s)

**Fig. 4: (a)** The projected magnetic moments along the *c*-axis in the case of an injection of one hole and one electron; **(b)** the band structures of the V-doped WS$_2$ monolayer with different doping carrier densities; and **(c)** the exchange energy and net magnetic moments of the V-doped WS$_2$ monolayer with different carrier doping densities. Net magnetic moment increases with increasing hole concentration, consistent with experimental results. A saturation feature is also present at higher hole concentrations, confirming what we observed experimentally.

**Figure 1**

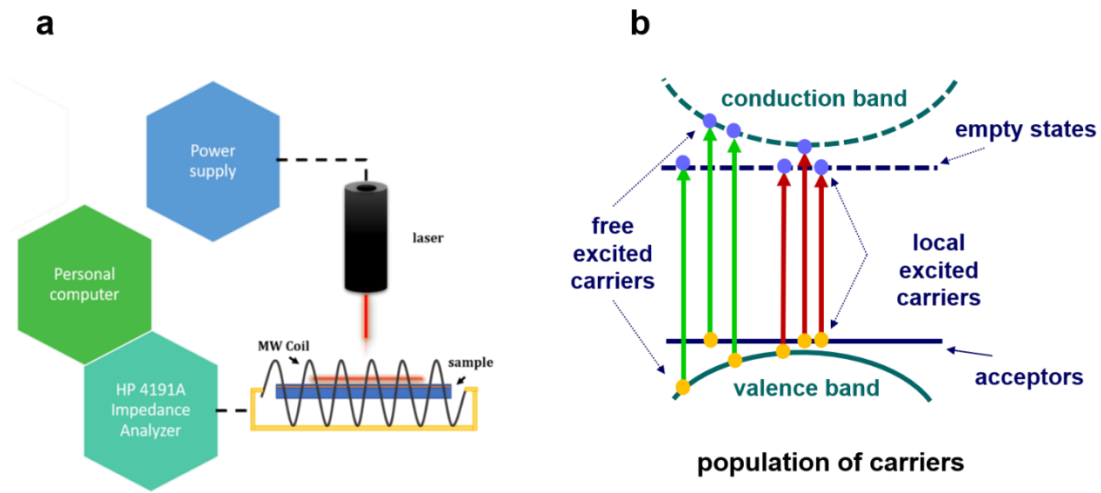

**Figure 2**

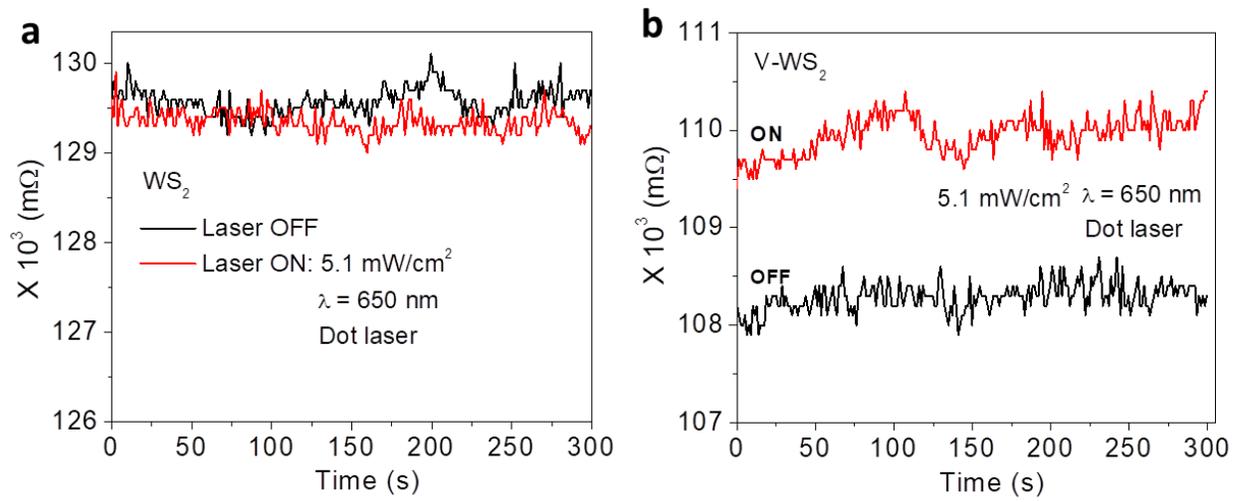

**Figure 3**

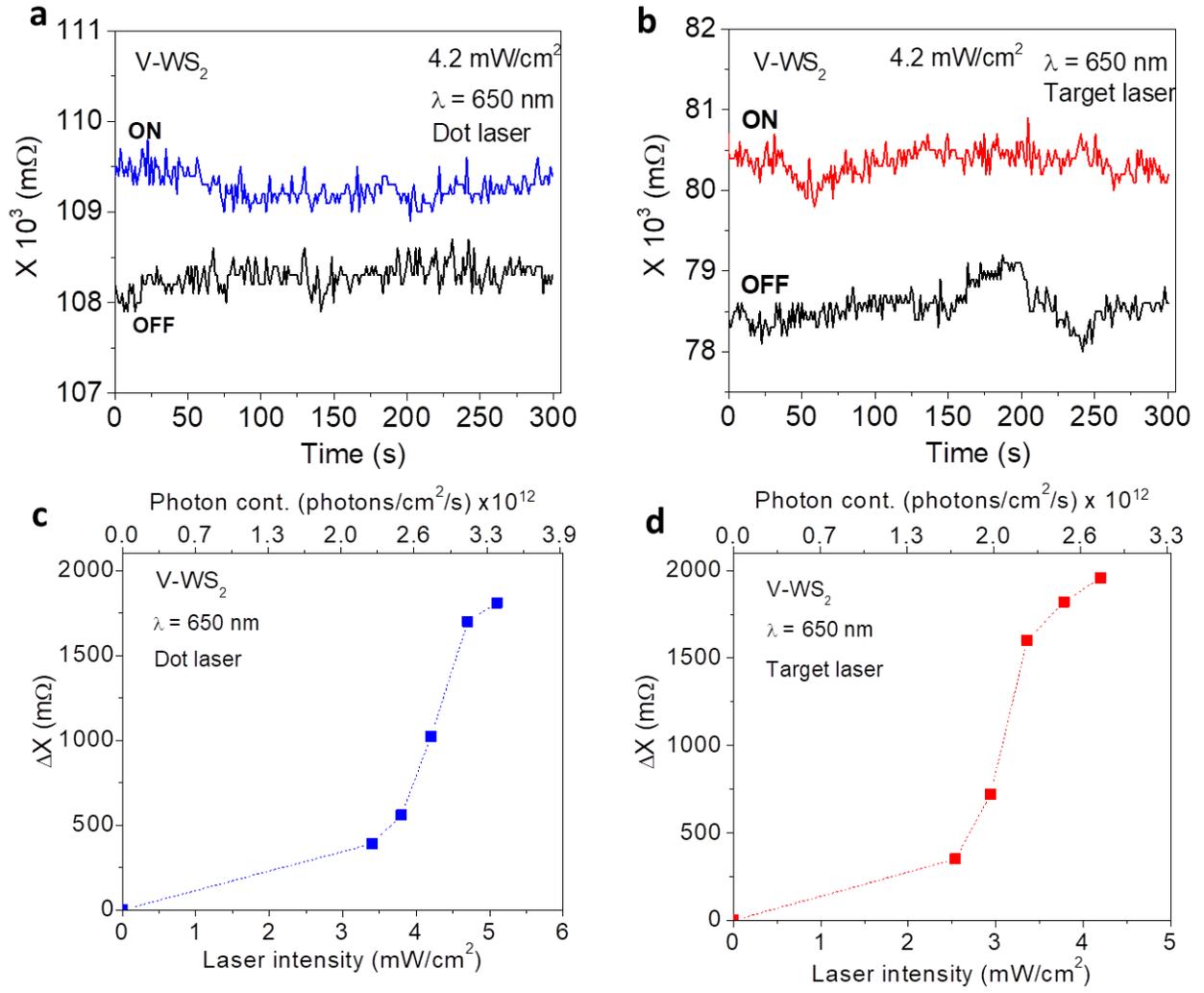

**Figure 4**

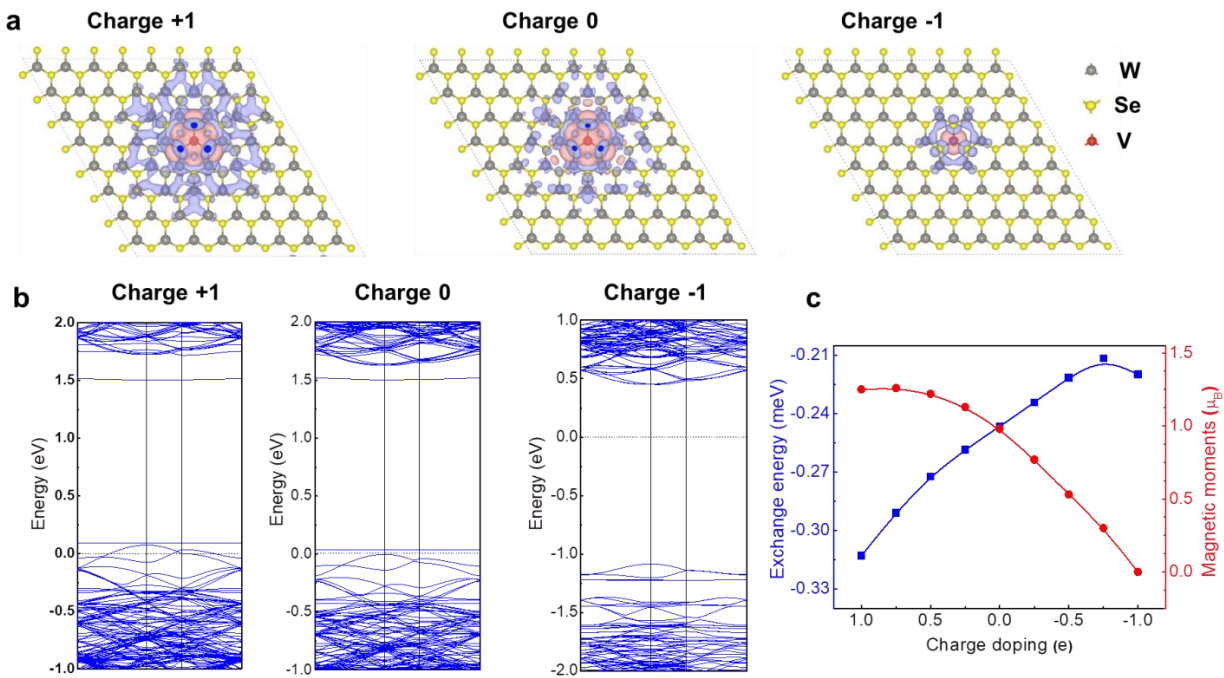